\documentclass{elsart1p}
 \usepackage{graphics}
\usepackage{amssymb}
\begin{document}
\begin{frontmatter}
%
%
%
%
%
\title{Percolation Approach to Initial Stage Effects in High Energy Collisions}
%
%

\author{Brijesh K Srivastava}

\address{Department of Physics, Purdue University, West Lafayette,
Indiana, USA}

\begin{abstract}
Possible phase transition of strongly interacting matter from hadron to a quark-gluon plasma state have in the past received considerable interest. The clustering of color sources provides a framework of the the partonic interactions in the initial stage of the collisions. The onset of de-confinement transition is identified by the spanning percolation cluster in 2D percolation. In this talk results are presented both for the multiplicity and the elliptic flow at RHIC and LHC energies. 
The thermodynamic quantities temperature, equation of state and transport coefficient are obtained in the framework of clustering of color sources. It is shown that the results are in excellent agreement with the recent lattice QCD calculations(LQCD).  

\end{abstract}

\begin{keyword}
%
Relativistic Heavy-Ion Collisions, Percolation, QGP, EOS
\PACS{12.38.Mh;} {25.75.Nq}
\end{keyword} 
\end{frontmatter}

\section{Introduction}
\label{}
One of the main goal of the study of relativistic heavy ion collisions is to study the deconfined matter, known as Quark-Gluon Plasma(QGP), which is expected to form at large densities. It has been suggested that the transition from hadronic to QGP state can be treated by percolation theory \cite{celik}. The formulation of percolation problem is concerned with elementary geometrical objects placed on a random d-dimensional lattice. Several object can form a cluster of communication. At certain density of the objects a infinite cluster appears which spans the entire system. This is defined by the dimensionless percolation density parameter $\xi$ \cite{isich}.  Percolation theory has been applied to several areas ranging from clustering in spin system to the formation of galaxies.
 In nuclear collisions there is indeed, as a function of parton density, a sudden onset of large scale color connection. There is a critical density at which the elemental objects form one large cluster, loosing their independent existence. Percolation would correspond to the onset of color deconfinement and it may be a prerequisite for any subsequent formation. Figure 1 shows the parton distribution in the transverse plane of a overlapping region of low and high density partons.

\begin{figure}
\centering        
\resizebox{0.70\textwidth}{!}{
\includegraphics{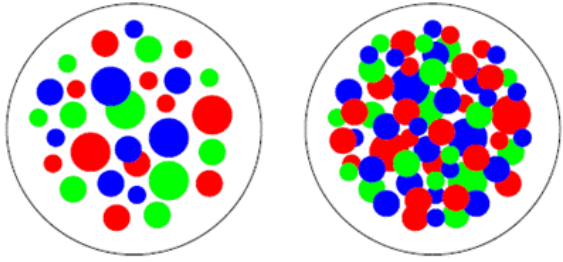}}
\caption{Partonic cluster structure in the transverse collision plane at low (left) and (right) high parton density \cite{satzextreme}.} 
\label{perc1}
\end{figure}

All high energy soft multi-hadron interactions exhibit thermal patterns of 
abundances characterized by the same temperature, independent of the center of mass energy \cite{braunmun,bec1}. The hadron limiting temperatures were measured by statistical thermal analyses that fit the data with a minimum of parameters \cite{braunmun,bec1}. In heavy ion collisions it may be plausible that multiple parton interactions produce a thermalized system.

 In this talk we present some of the results from the Color String Percolation Model (CSPM) e.g. for the multiplicity and  elliptic flow in A+A collisions. Results are also presented for the temperature, equation of state and the transport coefficient. 
 
\section{Clustering of Color Sources}   
Multi-particle production at high energies is currently described in terms of color strings stretched between the projectile and target. Hadronizing these strings produce the observed hadrons. At low energies only valence quarks of nucleons form strings that then hadronize. The number of strings grows with the energy and with the number of nucleons of participating nuclei. Color strings may be viewed as small discs in the transverse space filled with the color field created by colliding partons. Particles are produced by the Schwinger mechanisms \cite{schw}. With growing energy and size of the colliding nuclei the number of strings grow and start to overlap to form clusters \cite{pajares1,pajares2}. At a critical density a macroscopic cluster appears that marks the percolation phase transition. This is termed as Color String percolation Model (CSPM) \cite{pajares1,pajares2}. The interaction between strings occurs when they overlap and the general result, due to the SU(3) random summation of charges, is a reduction in the multiplicity and an increase in the string tension or an increase in the average transverse momentum squared, $\langle p_{t}^{2} \rangle$. 
We assume that a cluster of $\it n$ strings that occupies an area of $S_{n}$ behaves as a single color source with a higher color field $\vec{Q_{n}}$ corresponding to the vectorial sum of the color charges of each individual string $\vec{Q_{1}}$. The resulting color field covers the area of the cluster. As $\vec{Q_{n}} = \sum_{1}^{n}\vec{Q_{1}}$, and the individual string colors may be oriented in an arbitrary manner respective to each other , the average $\vec{Q_{1i}}\vec{Q_{1j}}$ is zero, and $\vec{Q_{n}^2} = n \vec{Q_{1}^2} $.

Knowing the color charge $\vec{Q_{n}}$ one can obtain the multiplicity $\mu$ and the mean transverse momentum squared $\langle p_{t}^{2} \rangle$ of the particles produced by a cluster of $\it n $ strings \cite{pajares2}
\begin{equation}
\mu_{n} = \sqrt {\frac {n S_{n}}{S_{1}}}\mu_{0};\hspace{5mm}
\langle p_{t}^{2} \rangle = \sqrt {\frac {n S_{1}}{S_{n}}} {\langle p_{t}^{2} \rangle_{1}}
\end{equation} 
where $\mu_{0}$ and $\langle p_{t}^{2}\rangle_{1}$ are the mean multiplicity and $\langle p_{t}^{2} \rangle$ of particles produced from a single string with a transverse area $S_{1} = \pi r_{0}^2$. In the thermodynamic limit, one obtains an analytic expression \cite{pajares1,pajares2}
\begin{equation}
\langle \frac {n S_{1}}{S_{n}} \rangle = \frac {\xi}{1-e^{-\xi}}\equiv \frac {1}{F(\xi)^2};\hspace{5mm}
F(\xi) = \sqrt {\frac {1-e^{-\xi}}{\xi}}
\end{equation}
where $F\xi)$ is the color suppression factor. $\xi = \frac {N_{s} S_{1}}{S_{N}}$ is the percolation density parameter assumed to be finite when both the number of strings $N_{S}$ and total interaction area $S_{N}$ are large.  Eq.(1) can be written as $\mu_{n}=F(\xi)\mu_{0}$ and 
$\langle p_{t}^{2}\rangle_{n} ={\langle p_{t}^{2} \rangle_{1}}/F(\xi)$.  
The critical cluster which spans $S_{N}$, appears for $\xi_{c} \ge$ 1.2 \cite{satz1}. 
It is worth noting that CSPM is a saturation model similar to the Color Glass Condensate (CGC), where $ {\langle p_{t}^{2} \rangle_{1}}/F(\xi)$ plays the same role as the saturation momentum scale $Q_{s}^{2}$ in the CGC model \cite{cgc,perx}. 
\section{Multiplicity in pp and A+A Collisions}

Measurements of particle multiplicities constrain the early time properties of colliding systems. In A+A case, these measurements are an essential ingredient for the estimation of the initial energy and entropy densities. The system will eventually thermalize to form the quark-gluon plasma. The charged particle multiplicity in A+A collisions at mid-rapidity is given by \cite{multi1}.  

\begin{equation}
\frac {1}{N_{A}} \frac {dn^{AA}}{dy} =\frac {dn^{pp}}{dy}\left ( 1+\frac{F(\xi)_{AA}}{F(\xi)_{pp}} (N_{A}^{\alpha(\sqrt s)}-1)\right )
\end{equation}
where $F(\xi)_{AA}$ and  $F(\xi)_{pp}$ are the color suppression factor for A+A and p-p collisions. $N_{A}$ is the average number of participating nucleons. $\alpha(\sqrt s)$ is a constant and in high energy limit it approaches 1/3 \cite{multi1}. 
\begin{figure}
\centering        
\vspace*{-1.cm}
\resizebox{0.75\textwidth}{!}{
\includegraphics{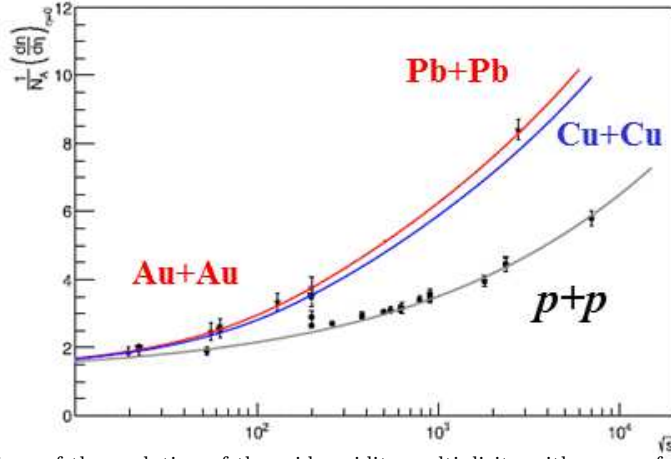}}
\vspace*{-0.5cm}
\caption{Comparison of the evolution of the mid-rapidity multiplicity with energy from the CSPM and data for pp and A+A collisions. Lines are from the model for pp (grey), Cu+Cu(blue) and red lines for Au+Au/Pb+Pb \cite{multi1}.} 
\label{perc2}
\end{figure}
Figure 2 shows a comparison of the evolution of the mid rapidity multiplicity with energy given by Eq.(3) with data for pp and A+A collisions.

\section{Elliptic Flow  $ v_{2}$}
The cluster formed by the strings has generally an asymmetric form in the transverse plane and acquires dimensions comparable to the nuclear overlap. This azimuthal asymmetry is at the origin of the elliptic flow in CSPM. The partons emitted at some point inside the cluster have to pass through the strong color field before appearing on the surface. Thus the energy loss by the parton is proportional to the length and therefore the $p_{t}$ of a particle will depend on the direction of the emission as shown in Fig.3. The percolation density parameter $\xi$ will be azimuthal angle dependent $\xi_{\phi} = \xi (R/{R_{\phi}})^{2}$. The $v_{2}$ expressed in terms of $\xi$ is given by \cite{flow1,flow2}

\begin{equation}
v_{2}=\frac {2}{\pi} \int^{\pi}_{0}d\phi cos(2\phi) \left (\frac{R_{\phi}}{R}\right)\left(\frac {e^{-\xi}-F(\xi)^{2}}{2F(\xi)^3}\right)\frac{R}{R-1}
\end{equation}

\begin{figure}
\centering        
\resizebox{0.45\textwidth}{!}{
\includegraphics{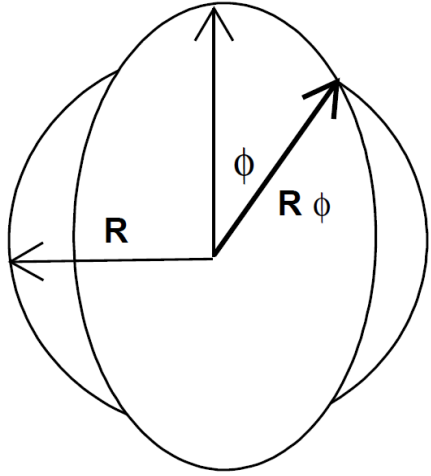}}
\caption{Azimuthal dependence of R. R being the radius of the projected circle \cite{flow1}.} 
\label{flowpic}
\end{figure} 
The transverse momentum dependence of $v_{2}$ computed using Eq.(4) for Pb+Pb at $\sqrt {s_{NN}}$=2.76 TeV and Au+Au  at $\sqrt {s_{NN}}$= 200 GeV is shown in Fig.4. The results are in good agreement with the ALICE and STAR results for $10-20\%$ centrality. 
\begin{figure}
\centering        
\resizebox{0.70\textwidth}{!}{
\includegraphics{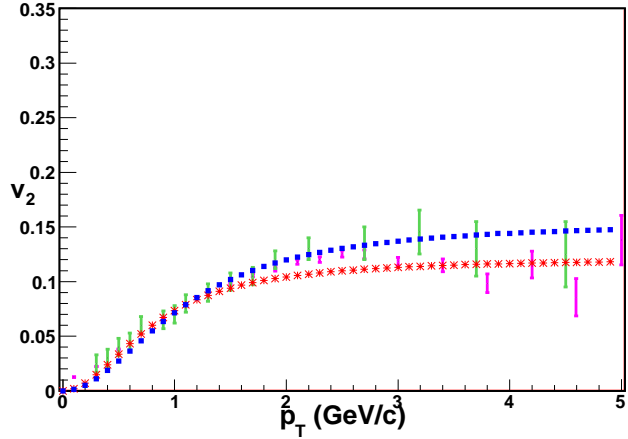}}
\caption{Elliptic flow $v_{2}$ comparison with data and CSPM. The error bars in red and green are the results from Pb+Pb at $\sqrt {s_{NN}}$=2.76 TeV and Au+Au  at $\sqrt {s_{NN}}$= 200 GeV. The CSPM results are shown as dotted blue and red lines \cite{flow2}.}
\label{flowv2}
\end{figure}
\section{Experimental Determination of  the Color Suppression Factor $F(\xi)$}
The suppression factor is determined by comparing the $\it pp$ and  A+A transverse momentum spectra. 
To evaluate the initial value of $F(\xi)$ from data for Au+Au collisions, a parameterization of $\it pp$ events at 200 GeV  is used to compute the $p_{t}$ distribution \cite{eos}
\begin{equation}
dN_{c}/dp_{t}^{2} = a/(p_{0}+p_{t})^{\alpha}
\end{equation}
where a is the normalization factor.  $p_{0}$ and $\alpha$ are parameters used to fit the data. This parameterization  also can be used for nucleus-nucleus collisions to take into account the interactions of the strings \cite{pajares2}
\begin{equation}
dN_{c}/dp_{t}^{2} = \frac {a'}{{(p_{0}{\sqrt {F(\xi_{pp})/F(\xi_{AA})}}+p_{t})}^{\alpha}}
\end{equation}
In pp collisions $F(\xi) \sim$ 1 at these energies due to the low overlap probability.$F(\xi)$ is related to $\xi$ by Eq.(2). 
\section{Temperature}
The connection between F($\xi$) and the temperature $T(\xi)$ involves the Schwinger mechanism (SM) for particle production. 
The Schwinger distribution for massless particles is expressed in terms of $p_{t}^{2}$ \cite{wong}
\begin{equation}
dn/d{p_{t}^{2}} \sim e^{-\pi p_{t}^{2}/x^{2}}
\end{equation}
where the average value of the string tension is  $\langle x^{2} \rangle$. The tension of the macroscopic cluster fluctuates around its mean value because the chromo-electric field is not constant.
The origin of the string fluctuation is related to the stochastic picture of 
the QCD vacuum. Since the average value of the color field strength must 
vanish, it can not be constant but changes randomly from point to point \cite{bialas}. Such fluctuations lead to a Gaussian distribution of the string tension, which transforms SM into the thermal distribution \cite{bialas}
\begin{equation}
dn/d{p_{t}^{2}} \sim e^{(-p_{t} \sqrt {\frac {2\pi}{\langle x^{2} \rangle}} )};\hspace{5mm}\langle x^{2} \rangle = \pi \langle p_{t}^{2} \rangle_{1}/F(\xi). 
\end{equation}
The temperature is expressed as \cite{eos,pajares3}  
\begin{equation}
T(\xi) =  {\sqrt {\frac {\langle p_{t}^{2}\rangle_{1}}{ 2 F(\xi)}}}
\end{equation} 

\section{Energy Density }
Among the most important and fundamental problems in finite-temperature QCD are the calculation of the bulk properties of hot QCD matter and characterization of the nature of the QCD phase transition. 
The QGP according to CSPM is born in local thermal equilibrium  because the temperature is determined at the string level. After the initial temperature $ T > T_{c}$ the  CSPM perfect fluid may expand according to Bjorken boost invariant 1D hydrodynamics \cite{bjorken}

\begin{equation}
\varepsilon = \frac {3}{2}\frac { {\frac {dN_{c}}{dy}}\langle m_{t}\rangle}{S_{n} \tau_{pro}}
\end{equation}
where $\varepsilon$ is the energy density, $S_{n}$ nuclear overlap area, and $\tau$ the proper time. Above the critical temperature only massless particles are present in CSPM. 
To evaluate $\varepsilon$ we use the charged pion multiplicity $dN_{c}/{dy}$ at midrapidity and $S_{n}$ values from STAR for 0-10\% central Au-Au collisions with $\sqrt{s_{NN}}=$200 GeV \cite{eos}. The factor 3/2 in Eq.(10) accounts for the neutral pions. The average transverse mass $\langle m_{t}\rangle$ is given by $\langle m_{t}\rangle =\sqrt {\langle p_{t}\rangle^2 + m_{0}^2} $, where $\langle p_{t}\rangle$ is the transverse momentum of pion and $m_{0}$ being the mass of pion.
\begin{equation}
\tau_{pro} = \frac {2.405\hbar}{\langle m_{t}\rangle}
\end{equation}
 In CSPM the total transverse energy is proportional to $\xi$. 
From the measured value of  $\xi$ and $\varepsilon$, as shown in Fig.5, 
it is found  that $\varepsilon$ is proportional to $\xi$ for the range 
$1.2 < \xi < 2.88$, $\varepsilon_{i}= 0.788$ $\xi$ GeV/$fm^{3}$ \cite{eos2}. This relationship has been extrapolated to below  $\xi= 1.2$ and  above $\xi =2.88$  for the energy density calculations normalized to $T^{4}$. Figure 6 shows $\varepsilon/T^{4}$ as obtained from CSPM along with the Lattice QCD results from HotQCD Collaboration \cite{bazavov}. 

\begin{figure}
\centering        
\resizebox{0.70\textwidth}{!}{
\includegraphics{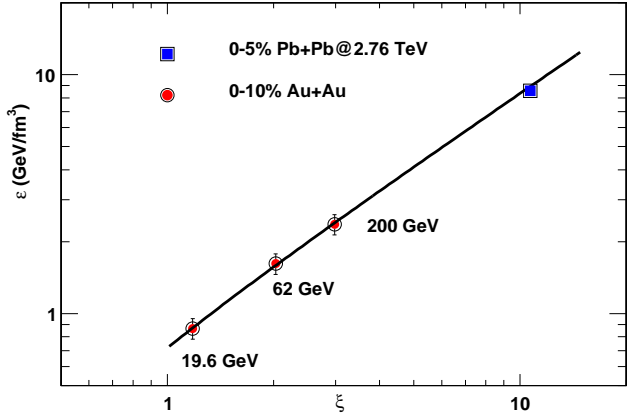}}
\caption{Energy density $\epsilon$ as a function of the percolation density parameter $\xi$. The extrapolated value for LHC energy is shown as blue square \cite{eos2}.} 
\label{perener}
\end{figure} 

\begin{figure}
\centering        
\resizebox{0.70\textwidth}{!}{
\includegraphics{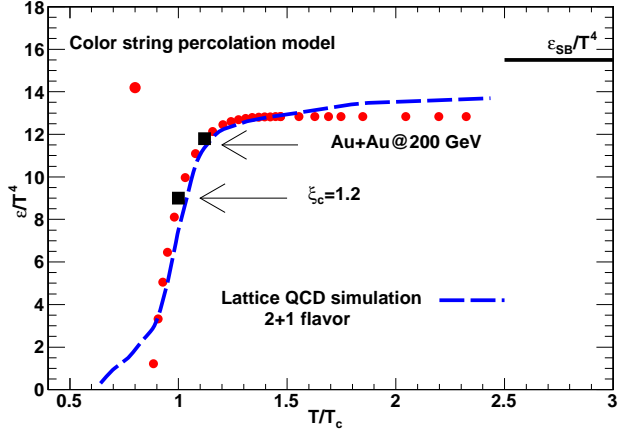}}
\caption{ $\epsilon/T^{4}$ versus $T/T_{c}$ from CSPM (red circles) and Lattice QCD (blue dash line) for 2+1 flavor and p4 action \cite{bazavov}.} 
\label{et4}
\end{figure} 
\section{Shear Viscosity}
 The relativistic kinetic theory relation for the shear viscosity over entropy density ratio, $\eta/s$ is given by \cite{gul1}
\begin{equation}
\frac {\eta}{s} \simeq \frac {T \lambda_{mfp}}{5}     
\end{equation}
where T is the temperature and $\lambda_{mfp}$ is the mean free path. $\lambda_{mfp} \sim \frac {1}{(n\sigma_{tr})}$ where 
$\it n $ is the number density of an ideal gas of quarks and gluons and $\sigma_{tr}$ the transport cross section. In CSPM the number density is given by the effective number of sources per unit volume \cite{eos2}
\begin{equation}
n = \frac {N_{sources}}{S_{N}L}
\end{equation}
 L is the longitudinal extension of the source, L = 1 $\it  fm $.
$\eta/s$ is obtained from $\xi$ and the temperature
\begin{equation}
\frac {\eta}{s} ={\frac {TL}{5(1-e^{-\xi})}} 
\end{equation}
\begin{figure}
\centering        
\resizebox{0.70\textwidth}{!}{
\includegraphics{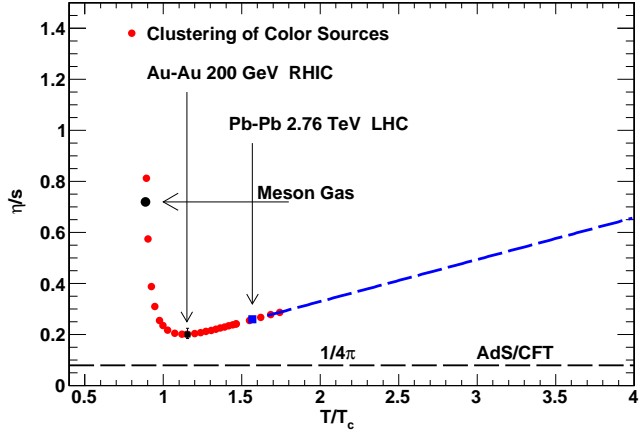}}
\caption{$\eta/s$ as a function of T/$T_{c}$. Au+Au at 200 GeV for 0-10$\%$ centrality is shown as solid black square.The estimated value for Pb+Pb at 2.76 TeV for 0-5$\%$ centrality  is shown as a solid blue square.The red dotted line represents the extrapolation to higher temperatures from the CSPM. The hadron gas value for $\eta/s$ $\sim$ 0.7 is shown as solid black circle at T/$T_{c} \sim $0.88 \cite{meson}.} 
\label{viscosity}
\end{figure} 
Figure 7 shows a plot of $\eta/s$ as a function of T/$T_{c}$. 
The lower bound shown in Fig. 7 is given by AdS/CFT \cite{kss}. The results from Au+Au at 200 GeV and Pb+Pb at 2.76 TeV collisions show that the $\eta/s$ value is 2.5 and 3.3 times the KSS bound \cite{kss}. 

\section{Trace Anomaly}
The trace anomaly ($\Delta$) is the expectation value of the trace of the energy-momentum tensor, $\langle \Theta_{\mu}^{\mu}\rangle = \varepsilon-3P$, which measures the deviation from conformal behavior and thus identifies the interaction still present in the medium. Both $\Delta$ and $\eta/s$ describe the transition from a strongly coupled QGP to a weakly coupled QGP. We find that the reciprocal of $\eta/s$ is in quantitative agreement with $(\varepsilon-3P)/T^{4}$, the trace anomaly over wide a range of temperature. This result is shown in Fig.8. The minimum in $\eta/s$ =0.20 at $T/T_{c}$= 1.15 determines the peak of the interaction measure $\sim$ 5 in agreement with the recent HotQCD values \cite{lattice12}. 
\begin{figure}
\centering        
\resizebox{0.70\textwidth}{!}{
\includegraphics{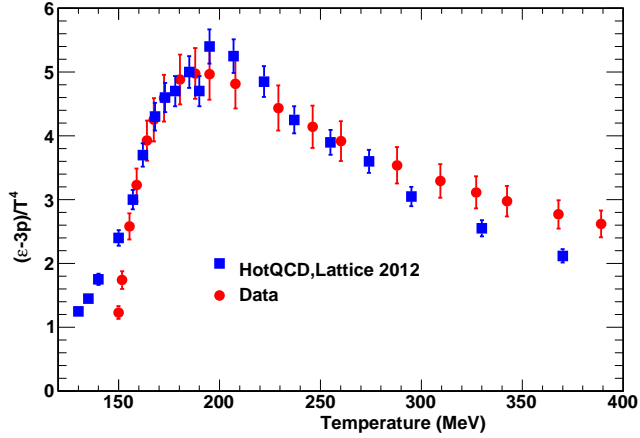}}
\caption{The trace anomaly $\Delta =(\varepsilon-3P)/T^{4}$ vs temperature \cite{lattice12}.} 
\label{trace}
\end{figure}
\section{Summary}
A brief review of the color string percolation model is presented. The 
clustering of color sources has described several observables in agreement with the experimental results e.g. multiplicity, elliptic flow etc. The model has been used to extract the color suppression factor from the experimental data. 
The thermodynamical quantities temperature and the equation of state are obtained in agreement with the LQCD calculation. The shear viscosity to entropy density ratio ($\eta/s$) are obtained at RHIC and LHC energies. It is also observed that the inverse of ($\eta/s$) is equivalent to trace anomaly $\Delta =(\varepsilon-3P)/T^{4}$. 

Thus the Clustering and percolation can provide a conceptual basis for the QCD phase diagram which is more general than the symmetry breaking \cite{satzx}.
 
\section{Acknowledgement}
This research was supported by the Office of Nuclear Physics within the U.S. Department of Energy  Office of Science under Grant No. DE-FG02-88ER40412. The author thanks A. Hirsch, C. Pajares and R. Scharenberg for fruitful discussions.

\end{document}